
\magnification=1200
\baselineskip=20pt
\def\lsim{<\kern-2.5ex\lower0.85ex\hbox{$\sim$}\ }
\def\rsim{>\kern-2.5ex\lower0.85ex\hbox{$\sim$}\ }
\overfullrule=0pt
\centerline{\bf TRIPLE PRODUCTS AND YANG--BAXTER EQUATION (I):}
\centerline{\bf OCTONIONIC AND QUATERNIONIC TRIPLE SYSTEMS}
\vskip 1cm
\centerline{by}
\vskip 1cm
\centerline{Susumu Okubo}
\centerline{Department of Physics and Astronomy}
\centerline{University of Rochester}
\centerline{Rochester, NY 14627, USA}
\vskip 3 cm
\noindent {\bf \underbar{Abstract}}

We can recast the Yang--Baxter equation as a triple  product equation.
Assuming the triple product to satisfy some algebraic relations, we can
find new solutions of the Yang--Baxter equation.  This program has been
completed here for the simplest triple systems which we call octonionic and
quaternionic.  The solutions are of rational type.
\vskip 2cm
\noindent {\underbar{PAC}: 03.65Fd. 02.90.+p
\vfil\eject
\noindent{\bf 1. \underbar{Introduction}}

The Yang--Baxter (Y--B) equation
$$\eqalign{&R^{b^\prime a^\prime}_{a_1b_1} (\theta)
R^{c^\prime a_2}_{a^\prime c_1} ( \theta^\prime)
R^{c_2 b_2}_{b^\prime c^\prime} (\theta^{\prime \prime})\cr
&= R^{c^\prime b^\prime}_{b_1c_1} (\theta^{\prime \prime})
R^{c_2 a^\prime}_{a_1 c^\prime} ( \theta^\prime)
R^{b_2 a_2}_{a^\prime b^\prime} (\theta)\cr}\eqno(1.1a)$$
with
$$\theta^\prime = \theta + \theta^{\prime \prime} \eqno(1.1b)$$
appears in many subjects ranging from statistical physics$^{1)}$,
 exactly solvable one-dimensional field theory$^{1),2)}$, and, the
braid group$^{3),4)}$ to the quantum group$^{1),5)}$.  Here, and hereafter,
all repeated indices are understood to be automatically summed over
$N$-values, $1,2,\dots, N$ for some integer $N$.
  It is often more convenient to consider a
 $N$-dimensional vector space $V$ over a field $F$ (normally real or
complex) with a basis vector $e_1, e_2, \dots, e_N$.  We ordinarily
introduce$^{1)}$ the general linear transformations $R_{12}(\theta),
R_{13} (\theta)$ and $R_{23} (\theta)$ operating on the tensor product
$V \otimes V \otimes V$ by
$$\eqalignno{R_{12} (\theta) e_a \otimes e_b \otimes e_c &=
R^{b^\prime a^\prime}_{ab} (\theta) e_{a^\prime} \otimes e_{b^\prime}
\otimes e_c &(1.2a)\cr
R_{23} (\theta) e_a \otimes e_b \otimes e_c &=
R^{c^\prime b^\prime}_{bc} (\theta) e_a \otimes e_{b^\prime}
\otimes e_{c^\prime} &(1.2b)\cr
R_{13} (\theta) e_a \otimes e_b \otimes e_c &=
R^{c^\prime a^\prime}_{ac} (\theta) e_{a^\prime} \otimes e_b
\otimes e_{c^\prime}\quad . &(1.2c)\cr}$$
Then, the Y--B equation Eq. (1.1a) is rewritten as
$$R_{12} (\theta) R_{13}(\theta^\prime) R_{23} (\theta^{\prime \prime}) =
 R_{23} (\theta^{\prime \prime}) R_{13} (\theta^\prime ) R_{12} (\theta)
\eqno(1.3)$$
which is convenient to discuss the underlying structure of the theory, such
as the Hopf algebra.  However,  it is not easy in general to find explicit
solutions of Eq. (1.3) except for some simple systems.

In this note, we will present a new way of solving the Y--B equation in
terms of some triple product systems$^{6),7)}$.  Here, we will first assume
for the sake of simplicity that $R^{ab}_{cd} (\theta)$ satisfies the
symmetric condition
$$R^{ba}_{dc} (\theta) = R^{ab}_{cd} (\theta) \eqno(1.4)$$
in what follows.
 The general case without assuming this
 will be discussed in the proceeding paper.
  Second, we suppose that the vector space $V$ possesses a
non-degenerate bilinear form $<x \vert y>\ :\ V \otimes V \rightarrow F$, and
set
$$g_{ab} =\  <e_a \vert e_b> \eqno(1.5)$$
with its inverse $g^{ab}$ so that
$$g^{ab} g_{bc} = \delta^a_c \quad . \eqno(1.6)$$
Although we ultimately restrict ourselves to the case of $g_{ab} =
 g_{ba}$ in this note, we need not yet assume it.  In terms of
$g^{ab}$, we set
$$e^a = g^{ab} e_b \quad , \quad e_a = g_{ab} e^b \eqno(1.7)$$
which satisfies
$$< e^a \vert e_b >\  = \delta^a_b \eqno(1.8)$$
as well as
$$e_a<e^a \vert x >\ = \ < x \vert e_a> e^a = x \quad . \eqno(1.9)$$

We now introduce a \underbar{$\theta$--dependent} triple product
$$[x,y,z]_\theta\  :\ V \otimes V \otimes V \rightarrow V\eqno(1.10)$$
by
$$\left[ e^b , e_c , e_d \right]_\theta
 = R^{ab}_{cd} (\theta) e_a \eqno(1.11)$$
or equivalently by
$$<e^a \vert \left[ e^b, e_c , e_d \right]_\theta> \
= R^{ab}_{cd} (\theta) \quad . \eqno(1.12)$$
The symmetric condition Eq. (1.4) can be readily rewritten as
$$< u \vert [z,x,y]_\theta>\ = \ <z \vert [u,y,x]_\theta> \quad ,
\eqno(1.13)$$
while the Y--B equation Eq. (1.1a) becomes a triple product equation
$$\eqalign{&[v,[u,e_j,z]_{\theta^\prime} \ , \
[e^j, x, y]_\theta ]_{\theta^{\prime \prime}}\cr
&\quad = [u,[v,e_j,x]_{\theta^\prime} \ , \
[e^j, z, y]_{\theta^{\prime \prime}} ]_\theta\cr}\eqno(1.14)$$
in a basis--independent notation.  Indeed, identifying $x = e_{a_1},\
y = e_{b_1},\ z = e_{c_1}, \ u = e^{a_2},$ and $v = e^{c_2}$ together with
Eq. (1.4), it will reproduce Eq. (1.1a).  We note also that Eq. (1.14) is
invariant under
$$u \leftrightarrow v \ ,\ x \leftrightarrow z \ , \
\theta \leftrightarrow \theta^{\prime \prime} \quad . \eqno(1.15)$$

Our ultimate task is to find solutions of the triple product equation
Eq. (1.14) under the condition Eq. (1.13).  To this end, we assume
existence of $\theta$--\underbar{independent} triple (or ternary) product
$$[x,y,z]\ :\ V \otimes V \otimes V \rightarrow V \eqno(1.16)$$
satisfying the symmetry condition
$$<u \vert [x,y,z]> \ = \ <z \vert [y, x,u]>\quad, \eqno(1.17)$$
and seek a solution of Eq. (1.14) in a form of
$$\eqalign{[z,x,y]_\theta = &P(\theta) [x,y,z] + A(\theta)
<x \vert y>z\cr
&+ B(\theta) <z \vert x>y + C(\theta) <z \vert y> x\cr}\eqno(1.18)$$
for some functions $P(\theta),\ A(\theta),\ B(\theta),$ and
$C(\theta)$ of $\theta$ to be determined.  Note first that we have changed
the orders of variables between both sides of Eq. (1.18) and second that the
condition Eq. (1.13) is automatically satisfied
 by Eq. (1.18), provided that we have
$$< y \vert x>\ =  \pm <x \vert y> \quad . \eqno(1.19)$$
Inserting the expression Eq. (1.18) to both sides of Eq. (1.14),
 and assuming
some suitable triple product relations for $[x,y,z],$ it will give a rather
complicated algebraic relations among $P(\theta),\ A(\theta), \
B(\theta),\ {\rm and}\ C(\theta)$, as we will see in sections 3 and 4 as
well as in subsequent papers.

In this paper, we will consider perhaps the simplest cases for the triple
product $[x,y,z]$ which we may call quaternionic and octonionic triple
products, where $[x,y,z]$ is totally anti-symmetric in 3 variables $x,\
y,\ {\rm and}\ z$.  We will give its general discussions in section 2, and
solve the corresponding Y--B equation in section 3 for $N = 8$ and in
section 4 for $N=4$.  The case of $N=8$ reproduces the previously known
result by de Vega and Nicolai$^{8)}$ who solved the Y--B equation in the
traditional component-wise fashion.  The more general triple systems will
be discussed in the subsequent paper$^{9)}$.  In the Appendix, we will
 explore some properties of octonionic triple systems, especially its
relation to the standard octonion algebra.
\smallskip
\noindent{\bf 2. \underbar{Octonionic and Quaternionic Triple Systems}}

In this section, we will present some mathematical analysis of triple
product systems which we call octonionic and quaternionic systems, since
they are needed for calculations of the next sections.  More details of
their mathematical structure will be found in the Appendix.  Let $V$ be the
$N$--dimensional vector space with non-degenerate inner product $<x
\vert y>$ and with the $\theta$--independent triple (or ternary) product
$[x,y,z]$ as in the previous section.  We assume now that they obey the
following axioms:
\item{(i)} $<x \vert y>\ = \ <y \vert x> \quad ,$\hfill (2.1)
\item{(ii)} $[x,y,z]$ is totally antisymmetric in $x,\ y,\ z$;\hfill (2.2)
\item{(iii)} $<w \vert [x,y,z]>$ is totally antisymmetric in
$x,\ y,\ z$ and $w$;\hfill (2.3)
\item{(iv)} $<[x,y,z] \vert [u,v,w]>$
\item{    } $= \alpha \{ <x \vert u><y \vert v><z \vert w> +
 <y \vert u><z \vert v><x \vert w>$
\item{    } $\quad + <z \vert u><x \vert v><y \vert w> -
\ <x \vert w><y \vert v><z \vert u>$
\item{    } $\quad - <y \vert w><z \vert v>
 <x \vert u> - <z \vert w><x \vert v><y \vert u>\}$
\item{    } $\quad + \beta\{ <x \vert u><y \vert [z,v,w]> +
 <y \vert u><z \vert [x,v,w]> $
\item{    } $\quad + <z \vert u><x \vert [y,v,w]>
+ <x \vert v><y \vert [z,w,u]>\ $
\vfill\eject
\item{    } $\quad + \ <y \vert v><z \vert
[x,w,u]>\ + \ <z \vert v><x \vert [y,w,u]>$
\item{    } $\quad + <x \vert w><y \vert [z,u,v]>\ + \ <y \vert w> <z \vert
[x,u,v]> $
\item{    } $\quad + \ <z \vert w><x \vert [y,u,v]>\}$ \hfill (2.4)

\noindent for some constants $\alpha$ and $\beta$.  In view of the
non-degeneracy of $<w \vert x>$ as well as by Eq. (2.3), the relation
Eq. (2.4) is equivalently rewritten as a triple product equation:
\item{(iv)$^\prime$} $[[x,y,z],u,v]$
\item{   } $= \{ \alpha [<y \vert v><z \vert u> \ - <y \vert u><z \vert
v>] - \beta <u \vert [v,y,z]>\}x$
\item{   } $\quad +\  \{ \alpha [<z \vert v><x \vert u>\ -\ <z \vert u><x
 \vert v>] - \beta <u \vert [v,z,x]>\}y$
\item{   } $\quad +\ \{ \alpha [<x \vert v><y \vert u> \ -\ <x
\vert u><y \vert v>] - \beta <u \vert [v,x,y]>\}z$
\item{   } $\quad - \ \beta \{<x \vert v>[u,y,z]\  + <y \vert v>[u,z,x]\  +
 \ <z \vert v>[u,x,y]$
\item{  } $\quad +
\ <x \vert u>[v,z,y]\  + \ <y \vert u>[v,x,z]\  +
\ <z \vert u>[v,y,x]\} \quad . \hfill (2.5)$

\noindent Before going into further details,
 we will prove shortly that the solutions
of Eqs. (2.1)--(2.5) are possible only for three cases of
\item{(a)} $N=8$ with $\alpha = \beta^2$ \hfill (2.6a)
\item{(b)} $N=4$ with $\beta =0$ \hfill (2.6b)
\item{(c)} $[[x,y,z],u,v] = 0$ identically
 with $\alpha = \beta = 0$. \hfill (2.6c)

\noindent Since the last possibility (c) is uninteresting, we will not
consider the case in this note.  We call then two cases of $N=8$ and $N=4$
to be octonionic and quaternionic triple systems, respectively, by a reason
to be explained in the Appendix.  We will also show there that the
underlying vector space $V$ for the case of $N=8$ can be identified as the
spinor representation space of the Lie algebra so(7).  Moreover, the reason
for the validity of Eq. (2.4) will also be given.

Let $e_1,e_2, \dots, e_N$ be a basis of $V$ normalized now by
$$<e_a \vert e_b>\  = \delta_{ab} \quad , \eqno(2.7)$$
and introduce the structure constant $c_{abcd}$ of the triple algebra by
$$[e_a, e_b, e_c] = c_{abcd} e^d \quad . \eqno(2.8)$$
Then, Eq. (2.3) implies that $c_{abcd}$ is totally antisymmetric in 4
indices $a,b,c,$ and $d$, while Eq. (2.4) is rewritten as
$$\eqalign{c_{abc \ell} c_{ijk}^{\ \ \ \ \ell} =\  &\alpha \sum_P (-1)^P
\delta_{ai} \delta_{bj} \delta_{ck}\cr
\noalign{\vskip 3pt}
&+ {1 \over 4}\ \beta \sum_P \sum_{P^\prime} (-1)^P (-1)^{P^\prime}
\delta_{ai} c_{bcjk} \quad , \cr}\eqno(2.9)$$
where the summations over $P$ and $P^\prime$ are over 3! permutations of $
a,b,c$ and of $i,j,k,$ respectively.  For $N=8$ with normalization
$\alpha = - \beta = 1$, Eq. (2.9) reproduces the result of de Wit and
 Nicolai$^{10)}$ as well as  that of G\"ursey and Tze$^{11)}$.
The case of $N=4$ is simpler.  Let $\epsilon_{ijk\ell}$ with
$\epsilon_{1234} = 1$ to be the totally antisymmetric Levi-Civita symbol in
4--dimensional space.  Then,
$$[e_i, e_j, e_k] = \epsilon_{ijk \ell} e^\ell \eqno(2.10)$$
defines the quaternionic triple system with $\alpha = 1$ and $\beta =0$, as
we will show in the Appendix.

Now, we will prove the results of Eqs. (2.6).  First, we calculate the
expression
$$\eqalign{J_0 &= [v,[u,e_j,z],[e^j,x,y]]\cr
&= -[v,[u,z,e_j],[x,y,e^j]]\cr}\eqno(2.11)$$
in the following way.  Setting
$$w = [u,e_j,z] = -[u,z,e_j] \quad , \eqno(2.12)$$
we evaluate
$$\eqalign{J_0 &= -[[x,y,e^j],w,v]\cr
&= -\{ \alpha [<y \vert v><e^j \vert w>\ -\
<y \vert w><e^j \vert v>] - \beta <w \vert
[v,y,e^j]>\}x\cr
&\quad -\{ \alpha [<e^j \vert v><x \vert w>\ -\ <e^j \vert w><x \vert v>]
- \beta <w \vert [v,e^j,x]>\}y\cr
&\quad -\{ \alpha [<x \vert v><y \vert w>\ - \ <x \vert w>
<y \vert v>] - \beta <w \vert [v, x, y]>\} e^j\cr
&\quad + \beta \{<x \vert v>[w,y,e^j]\  + \ <y \vert
v>[w,e^j,x]\  +\ <e^j \vert v> [w,x,y]\cr
&\quad + \ <x \vert w>[v,e^j,y]\  + <y \vert w> [v,x,e^j]\  + \
<e^j \vert w>[v,y,x]\}\cr}\eqno(2.13)$$
by replacing $z$ and $u$ by $e^j$ and $w$, respectively, in Eq. (2.5).  We
calculate for example
$$\eqalign{[w,e^j,x] &= -[[u,z,e_j],e^j,x]\cr
&= \alpha (N-2)\{<x \vert u>z - \ <x \vert z>u\}
- \beta (N-4)\ [x,u,z]\cr}$$
again from Eq. (2.5) together with Eq. (1.9).
  Also, we can reduce a term such as
$$<w \vert [e^j, x, y]>\ = \ <[u,e_j, z]\vert
[e^j, x, y]>$$
into a simpler form when we utilize Eq. (2.4).  Inserting these results
into (2.13) and noting Eq. (1.9), we find
$$\eqalign{J_0 &= -[v,[u,z,e_j],[x,y,e^j]]\cr
&= A_1 x + \overline{A_1} y + A_2 u + \overline{A_2} z\cr
&\quad + \{ (N-6) \beta^2 - \alpha \} \{ <x \vert v>
[y,u,z]\  - \ <y \vert v>[x,u,z]\}\cr
&\quad + \beta^2 \{ -<x \vert u>[y,z,v]\  +\
<y \vert u>[x,z,v]\  + <x \vert z>[y,u,v]\cr
&\quad -<y \vert z>[x,u,v]\  -\ <u \vert v>[x,y,z]\  +
\ <z \vert v> [x,y,u]\}\cr}\eqno(2.14)$$
after some calculations where we have set
$$\eqalignno{A_1 &= -\{ (N-6) \beta^2 - \alpha\} <y \vert [z,u,v]>\cr
&\quad + \alpha \beta (N-4) \{<y \vert u><z \vert v>\ -\ <y \vert z><v
 \vert u>\}\quad , &(2.15a)\cr
\overline{A_1} &= \{ (N-6) \beta^2 - \alpha\} <x \vert [u,v,z]>\cr
&\quad - \alpha \beta (N-4) \{<x \vert u><z \vert v>\ -\ <x \vert z><v
 \vert u>\}\quad , &(2.15b)\cr
A_2 &= \beta^2 <y \vert [x,z,v]>\cr
&\quad + \
\alpha \beta (N-4)\{<x \vert v><y \vert z>\ -
\ <y \vert v><x \vert z>\}\quad , &(2.15c)\cr
\overline{A_2} &= -\beta^2 <y \vert [x,u,v]>\cr
 &-\  \alpha \beta (N-4) \{
<x \vert v><y \vert u> \ -\ <y \vert v><x \vert u>\} \quad . &(2.15d)\cr}$$
However, the left side of Eq. (2.14) is antisymmetric for the exchange of
$x \leftrightarrow u$ and $y \leftrightarrow z$, so that the right side
of Eq. (2.14) must share the same property.  This fact can be easily seen
to lead to the validity of either
$$(N - 7) \beta^2 = \alpha \eqno(2.16)$$
or
$$\eqalign{<&x \vert v>[y,u,z]\  -
\ <y\vert v>[x,u,z]\  + \
<u \vert v>[x,y,z]\  -\
<z \vert v>[x,y,u]\cr
&= \ <y \vert [z,u,v]>x\  - \ <x \vert [z,u,v]>y\cr
&\quad +\
<y \vert [x,z,v]>u\  -\
<y \vert [x,u,v]> z \quad .\cr}\eqno(2.17)$$
For the second case of Eq. (2.17), we set further $u = e_j$ and
$v = e^j$ and sum over $j=1,2,\dots,N$ to find
$$(N -4)[x,y,z] = 0 \quad . \eqno(2.17)^\prime$$
Therefore, excluding the trivial case of $[x,y,z] = 0$ identically, the
second condition Eq. (2.17) is possible only for $N=4$.

Next, we calculate another expression
$$J_1 = [[x,y,z],e_j,[u,v,e^j]]$$
in the following two different ways.  First, we set $w = [u,v,e^j]$ and
evaluate
$$J_1 = [[x,y,z],e_j,w]$$
by repeated uses of Eq. (2.5).  Alternately, we can compute $J_1$
 in a similar way by
setting $\overline w = [x,y,z]$ and hence
$$J_1 = [ \overline w, e_j, [u,v,e^j]] =
 [[u,v,e^j],\overline w, e_j]\quad .$$
Comparing both expressions, we obtain
$$\eqalign{2(&\beta^2 - \alpha)\{<y \vert [u,v,z]>x\  +
\ <z \vert [u,v,x]>y\  + \ <x \vert [u,v,y]>z\}\cr
&+ [6 \beta^2 - (N-2) \alpha]\{ <v \vert [x,y,z]>u\  -\ <u \vert
[x,y,z]>v\}\cr
&- (N-8)\beta^2 \{ <x \vert u> [y,z,v]\  +\
<y \vert u>[z,x,v]\  + \ <z \vert u>[x,y,v]\cr
&- \ <x \vert v>[z,y,u]\  -\
<y \vert v>[x,z,u]\  -\ <z \vert v>[y,x,u]\} =0 \quad .\cr}\eqno(2.18)$$
Setting $u=e_j$ and $v = e^j,$ and summing over $j$, Eq. (2.18) leads to
$$-6 \beta^2 (N-8) [x,y,z] = 0$$
so that we have $N=8$ or $\beta=0$, provided that $[x,y,z]$ is not
identically zero.  Combining this fact with Eqs. (2.16) and
(2.17)$^\prime$,
 we find the desired  results Eqs. (2.6).  Note that for $N=8$ with $\alpha =
 \beta^2$, Eq. (2.18) is identically satisfied, while, for $N=4$ with
$\beta = 0$, it gives an identity
$$\eqalign{<&y \vert [u,v,z]>x\  + \ <z \vert [u,v,x]>y\  +
 \ <x \vert [u,v,y]>z\cr
&+ <v \vert [x,y,z]>u\  -\ <u \vert [x,y,z]>v = 0 \quad .\cr}
\eqno(2.19)$$
The relation Eq. (2.19) is automatically satisfied for $N=4$, since its
left side is totally antisymmetric in 5 variables $x,y,z,u,$ and $v$.
Relations Eqs. (2.17) and (2.19) will be utilized in section 4.

For the case of $N=8$ with $\alpha = \beta^2$, Eq. (2.14) is rewritten as
$$\eqalignno{J_0 &= [v,[u,e_j,z],[e^j,x,y]]\cr
&= \{4 \alpha \beta [<y \vert u><z \vert v> \ -\ <y \vert z><v \vert
u>] - \beta^2 <y \vert [z,u,v]>\}x\cr
&\quad -\{ 4 \alpha \beta [<x \vert u><z \vert v> \ -\ <x \vert z><v \vert
u>] - \beta^2 <x \vert [u,v,z]>\}y\cr
&\quad + \{ 4 \alpha \beta [<x \vert v><y \vert z>\ -\ <y \vert v><x \vert
z>] + \beta^2 <y \vert [x,z,v]>\}u &(2.20)\cr
&\quad - \{ 4 \alpha \beta [<x \vert v><y \vert u>\ -\ <y \vert v><x \vert
u>] + \beta^2 <y \vert [x,u,v]>\}z\cr
&\quad + \beta^2 \{ <x \vert v>[y,u,z]\   -\ <y \vert v>[x,u,z]\  -
<x \vert u> [y,z,v]\   + <y \vert u>[x,z,v]\cr
&\quad + <x \vert z>[y,u,v]\ -\ <y \vert z>[x,u,v]\  -
<u \vert v> [x,y,z] + <z \vert v>[x,y,u]\}\cr}$$
which will be used in the calculation of the next section.

We can similarly compute
$$[[x,y,z],t,[u,v,w]]$$
in two different ways.  In this way, we obtain a very complicated identity
which is however not given here.

Further discussions of the triple product as well as its relation to the
octonion and quaternion algebras will be given in the Appendix.

\smallskip
\noindent {\bf 3. \underbar{Solution for $N=8$}}

Here, let $[x,y,z]$ be the octonionic triple product as has been defined in
the previous section.  We seek a solution of the Y--B equation Eq. (1.14)
in a form given by Eq. (1.18) for 4--unknown functions $P(\theta),\
A(\theta),\ B(\theta), \ {\rm and}\ C(\theta)$ to be determined.  For
simplicity, we set
$$P = P(\theta)\quad , \quad P^\prime = P(\theta^\prime) \quad , \quad
P^{\prime \prime} = P(\theta^{\prime \prime}) \quad , \eqno(3.1)$$
and similarly for $A(\theta),\ B(\theta), \ {\rm and}\ C(\theta)$.
Inserting the expression Eq. (1.18) into Eq. (1.14), each side of Eq.
(1.14) contains 64 terms,  among whom the most complicated term is the one
proportional to $P^{\prime \prime} P^\prime P$, i.e.
$$P^{\prime \prime} P^\prime P \{[v,[u,e_j,z],
[e^j,x,y]] - (u \leftrightarrow v, x \leftrightarrow z)\} \quad .
\eqno(3.2)$$
However, for $N=8$, this can be computed easily from Eq. (2.20).  At any
rate, after some straightforward calculations, we find the following
result:
$$\eqalign{0 &= [v,[u,e_j,z]_{\theta^\prime},
[e^j,x,y]_\theta]_{\theta^{\prime \prime}}
 - (u \leftrightarrow v,x \leftrightarrow z, \theta
\leftrightarrow \theta^{\prime \prime})\cr
&= G_1 <x \vert z>[y,u,v] + G_2 <u \vert v> [x,y,z] +
G_3 <x \vert u> [z,y,v]\cr
&\quad - \hat G_3 <z \vert v>[x,y,u]
+  G_4 <x \vert v>[y,z,u] - \hat  G_4 <z \vert u> [y,x,v]\cr
&\quad + G_5 <y \vert z> [x,u,v] - \hat G_5 <y \vert x>[z,v,u]
+ G_6 <y \vert u>[x,z,v]\cr
&\quad - \hat G_6 <y \vert v> [z,x,v] +
G_7 <z \vert u> <y \vert v>x -
\hat G_7 <x \vert v><y \vert u>z\cr
&\quad + G_8 <y \vert z><v \vert u>x - \hat G_8 <y \vert x>
<u \vert v>z  +
G_9 <x \vert v><y \vert z>u\cr
&\quad - \hat G_9 <z \vert u><y \vert x>v
+ G_{10} <y \vert v><x \vert z>u - \hat G_{10} <y \vert u>
<z \vert x> v\cr
&\quad + G_{11} <y \vert [z,u,v]>x  -
\hat G_{11} <y \vert[x,v,u]>z \cr
&\quad + G_{12} <y \vert [x,z,v]>u - \hat G_{12} <y \vert [z,x,u]>v\cr}
\eqno(3.3)$$
where $G_\mu (\mu = 1,2,\dots,12)$ are cubic polynomials of $P,\ A,\ B,\
{\rm and}\ C$ to be given below, and $\hat G_\mu$ is the function which can
be obtained from $G_\mu$ by interchanging $\theta \leftrightarrow
\theta^{\prime \prime}$.  Their explicit forms are found to be
$$\eqalign{G_1 &= 2 \beta^2 P^{\prime \prime} P^\prime P - 2 \beta
 P^{\prime \prime} B^\prime P - \beta P^{\prime \prime} P^\prime B -
\beta B^{\prime \prime} P^\prime P - P^{\prime \prime} A^\prime
B - B^{\prime \prime} A^\prime P \quad ,\cr
G_2 &= -G_1 \quad ,\cr
G_3 &= 0 \quad ,\cr
G_4 &= \beta^2 P^{\prime \prime} P^\prime P + 2 \beta P^{\prime \prime}
B^\prime P + \beta P^{\prime \prime} P^\prime B + \beta B^{\prime \prime}
P^\prime P \cr
&\quad -C^{\prime \prime} P^\prime C - B^{\prime \prime}C^\prime P +
C^{\prime \prime} C^\prime P - P^{\prime \prime} C^\prime B +
 P^{\prime \prime} C^\prime C + 4 \beta P^{\prime \prime}
C^\prime P \quad ,\cr
G_5 &= -\beta^2 P^{\prime \prime} P^\prime P +  \beta P^{\prime \prime}
B^\prime P + 4 \beta A^{\prime \prime} P^\prime P - \beta P^{\prime \prime}
P^\prime B \cr
&\quad + 2 \beta B^{\prime \prime} P^\prime P - A^{\prime \prime} P^\prime
 B  + A^{\prime \prime} P^\prime C - A^{\prime \prime} A^\prime P +
 A^{\prime \prime} B^\prime P - P^{\prime \prime}
A^\prime C \quad ,\cr
G_6 &= 2 \beta^2 P^{\prime \prime} P^\prime P + \beta P^{\prime \prime}
B^\prime P - \beta B^{\prime \prime} P^\prime P -  P^{\prime \prime}
B^\prime C + 2 \beta P^{\prime \prime} P^\prime B + B^{\prime \prime}
P^\prime C \quad ,\cr
G_7 &= 4 \alpha \beta P^{\prime \prime} P^\prime P
 - \alpha P^{\prime \prime}
B^\prime P + \alpha P^{\prime \prime} P^\prime B + \alpha
 B^{\prime \prime}
P^\prime P \cr
&\quad - 6 \alpha P^{\prime \prime} C^\prime P + B^{\prime \prime}
C^\prime C +
C^{\prime \prime} C^\prime B - C^{\prime \prime} B^\prime C
\quad ,\cr
G_8 &= - 4 \alpha \beta P^{\prime \prime} P^\prime P
+ \alpha  P^{\prime \prime}
B^\prime P - 6 \alpha  A^{\prime \prime} P^\prime P -
\alpha P^{\prime \prime}
P^\prime B \cr
&\quad - \alpha B^{\prime \prime} P^\prime P + B^{\prime \prime}
A^\prime C -
A^{\prime \prime} A^\prime B - A^{\prime \prime} B^\prime C \quad ,\cr
G_9 &= 4 \alpha \beta P^{\prime \prime} P^\prime P
- \alpha  P^{\prime \prime}
B^\prime P + \alpha  P^{\prime \prime} P^\prime B +
\alpha B^{\prime \prime}
P^\prime P - A^{\prime \prime} A^\prime A
- A^{\prime \prime} B^\prime A \cr
&\quad - 8 A^{\prime \prime} C^\prime A - B^{\prime \prime}
C^\prime A -
C^{\prime \prime} C^\prime A - A^{\prime \prime} C^\prime B
 - A^{\prime \prime} C^\prime C + C^{\prime \prime} A^\prime C \quad ,\cr
G_{10} &= \hat G_8 \quad ,\cr
G_{11} &= - G_6 \quad ,\cr
G_{12} &= - \hat G_5 \quad .\cr}\eqno(3.4)$$
Therefore, the Y--B equation is satisfied, if we have
$$G_1 = G_4 = G_5 = G_6 = G_7 = G_8 = G_9 = 0 \quad . \eqno(3.5)$$
The conditions Eq. (3.5) agree with those found by de Vega and
Nicolai$^{8)}$, if we set $\alpha = - \beta = 1$, and
$\theta \rightarrow \theta^{\prime\prime} \rightarrow \theta^\prime
\rightarrow \theta$ with $P(\theta) = - c (\theta),\ A(\theta) = b
(\theta),\ {\rm and}\ B(\theta) = a (\theta)$.  For simplicity, we
normalize $\alpha$ and $\beta$ by
$$\alpha = - \beta = 1 \quad . \eqno(3.6)$$
Then, the solution of Eq. (3.5) can be found by first considering
$G_1 = 0$ which can be rewritten as
$$\left( {B^{\prime \prime} \over P^{\prime\prime}} +
{B \over P}\right)\left( {A^\prime \over P^\prime} - 1 \right)
 = 2 + 2 \ {B^\prime \over P^\prime} \quad . \eqno(3.7)$$
However, since we must have $\theta^\prime = \theta + \theta^{\prime
\prime}$, the solution must be of form
$${B \over P} = {B (\theta) \over P(\theta)} =\lambda_0 + b \theta
\eqno(3.8)$$
for some constants $\lambda_0$ and $b$, so that
$${A \over P} = {A (\theta) \over P (\theta)} = 1 +
 {2 (1 + \lambda_0 + b \theta) \over
2 \lambda_0 + b \theta} \quad .$$
We can then examine the rest of Eqs. (3.5) to find the final solution:
$$\eqalignno{{A (\theta) \over P (\theta)} &=
{18 - 3 b \theta \over 10 - b \theta} \quad , &(3.9a)\cr
\noalign{\vskip 3pt}
{B (\theta) \over P (\theta)} &= b \theta - 5 \quad , &(3.9b)\cr
\noalign{\vskip 3pt}
{C (\theta) \over P (\theta)} &= {12 - 3 b \theta
 \over b \theta} \quad ,&(3.9c)\cr}$$
where $b$ is an arbitrary constant and $P(\theta)$ is undetermined.  With
the choice of normalization $b= -3$ and $C(\theta) = {1 \over
\theta}$, it reproduces the result of reference 8,
$$\eqalignno{P(\theta) &= - {1 \over 3 \theta +4} \quad , &(3.10a)\cr
\noalign{\vskip 3pt}
A(\theta) &= - {9 (\theta + 2) \over
(3 \theta +4)(3 \theta + 10)}\quad , &(3.10b)\cr
\noalign{\vskip 3pt}
B(\theta) &= {3 \theta +5 \over 3 \theta +4} \quad , &(3.10c)\cr
\noalign{\vskip 3pt}
C(\theta) &= {1 \over \theta} \quad . &(3.10d)\cr}$$

Although we have assumed $P(\theta) \not= 0$ in our derivation, we can find
a solution for $P(\theta) = 0$ by first setting $P(\theta) = b/12$ in Eqs.
(3.9) and then letting $b \rightarrow 0$.  In this way, we obtain a rather
trivial classical Y--B  solution
$$\eqalign{P(\theta) &= A(\theta) = B(\theta) = 0 \quad ,\cr
C(\theta) &= {1 \over \theta} \quad .\cr}\eqno(3.11)$$
The more general solution for $P(\theta) = 0$ will be given in the
proceeding paper.
\medskip
\noindent {\bf 4. \underbar{Solution for $N=4$}}

As we will explain shortly, the case of $N=4$ requires a separate
discussion.  Surprisingly, this case is slightly more complicated than the
previous one of $N=8$, although it is simpler for calcualtions of triple
products.  For instance, since we have $\beta = 0$, Eq. (2.14) reduces to a
simpler equation
$$\eqalign{[v,&[u,e_j,z],[e^j,x,y]]\cr
&= \alpha <y \vert [z,u,v]>x - \alpha <x\vert [u,v,z]>y\cr
&\quad - \alpha \{ <x \vert v>[y,u,z]\  - <y \vert v>[x,u,z]\} \quad .
\cr}\eqno(4.1)$$
However, we have, now, various constraints to be taken into account.  For
any 6 variables $x,y,z,u,v,$ and $w$, we first note the validity of the
identity
$$\eqalign{<w\vert x><y \vert [z,u,v]>\ &+ \ <w \vert y><z \vert
[u,v,x]>\ + \ <w \vert z><u \vert [v,x,y]>\cr
&+\ <w \vert u><v \vert [x,y,z]>\ + \
<w \vert v><x\vert [y,z,u]>\ = 0\cr}\eqno(4.2)$$
for $N=4$, since the left side of Eq. (4.2) is totally antisymmetric in 5
variables $x,y,z,u$ and $v$.  In view of the non-degeneracy of the inner
product $<w \vert x>$, Eq. (4.2) immediately give
$$\eqalign{<y\vert [z,u,v]>x\  &+ \ <z\vert [u,v,x]> y\  + <u \vert
[v,x,y]>z\cr
&+ \ <v \vert [x,y,z]>u\  + <x \vert [y,z,u]>v = 0 \cr}\eqno(4.3)$$
which is equivalent to Eq. (2.19).  Moreover, if we note
$$\eqalign{<y \vert [z,u,v]>\ &= - <z\vert [y,u,v]>\cr
<v \vert [x,y,z]> \ &= -<z \vert [x,y,v]>\cr
<x \vert [y,z,u]>\ &= -<z \vert [y,x,u]> \quad ,\cr}$$
then Eq. (4.2) leads also to another identity
$$\eqalign{&- <w \vert x>[y,u,v]\  + <w \vert y> [u,v,x]\  +\  <u \vert
[v,x,y]>w\cr
&- <w \vert u>[x,y,v]\  - \ <w \vert v> [y,x,u] = 0\quad .\cr}$$
Rewriting $w$ by $z$, and noting Eq. (4.3), we find then
$$\eqalignno{<z &\vert y>[u,v,x]\cr
&=\  <z \vert x>[y,u,v]\  + <z \vert u>[x,y,v]
\ + \ <z \vert v>[y,x,u]\  - \ <u \vert [v,x,y]>z\cr
&=\ <z \vert x>[y,u,v]\  + \ <z \vert u>[x,y,v]\  +
\ <z \vert v>[y,x,u]&(4.4)\cr
&\quad -\ <z \vert [y,u,v]>x\  + \
<z \vert [u,v,x]>y\  - \ <z \vert [x,y,v]>u\  +
\ <z\vert [x,y,u]>v \cr}$$
which reproduces Eq. (2.17) when we let $z \leftrightarrow v$.
Interchanging $z \leftrightarrow x,\ {\rm or}\ z \leftrightarrow u,\
{\rm or}\ z
\leftrightarrow~v$ etc., we can therefore eliminate now
 $\langle z \vert y \rangle [u,v,x], \
\langle x \vert y \rangle [u,v,z],\ \langle u \vert y
\rangle [z,v,x]\ {\rm and}\ \langle v~\vert y \rangle [u,z,x]$
as linear combinations of terms of form $\langle a \vert b
 \rangle [c,d,y],\ \langle y \vert
[a,b,c]\rangle d \ {\rm and}\  \langle z~\vert [u,v,x]
\rangle y \ {\rm for}\ a,b,c,d$ being some
permutations of $x,z,u\ {\rm and}\ v$.  Except for this fact, the
calculation for $N=4$ proceeds just as in the previous case of $N=8$ to
lead to
$$\eqalign{0 &= [v,[u,e_j,z]_{\theta^\prime},
[e^j,x,y]_\theta]_{\theta^{\prime \prime}} - (u \leftrightarrow v,
x \leftrightarrow z, \theta \leftrightarrow \theta^{\prime \prime})\cr
&= F_1 <u \vert v>[z,x,y] + F_2 <x \vert z>[y,u,v] +
F_3 <x \vert u>[v,z,y] \cr
&\quad - \hat F_3 <z \vert v>[u,x,y] + F_4 <x \vert v>[y,z,u] - \hat F_4
<z \vert u>[y,x,v]\cr
&\quad + F_5 <y \vert [z,u,v]>x - \hat F_5 <y \vert
[x,v,u]>z + F_6 <y \vert [z,x,v]>u\cr
&\quad - \hat F_6 <y \vert [z,x,u]>v + F_7 <y \vert z>
<u \vert v>x - \hat  F_7 <y \vert x><u \vert v>z \cr
&\quad + F_8 <z \vert u><y \vert v>x  - \hat F_8 <x \vert v><y \vert u> z
+ F_9 <x \vert z><y \vert v>u \cr
&\quad - \hat F_9 <x \vert z><y \vert u>v  + F_{10} <y \vert z>
<x \vert v>u
- \hat F_{10} <y \vert x><z \vert u> v \quad ,\cr}\eqno(4.5)$$
where $\hat F_\mu (\mu = 1,2,\dots , 10)$ is the function obtained from
$F_\mu$ by interchanging $\theta \leftrightarrow \theta^{\prime\prime}$ as
before.  Explicit expressions of $F_\mu$'s are found after some
calculations to be
$$\eqalign{F_1 &= 2 \alpha P^{\prime \prime}P^\prime P - C^{\prime\prime}
P^\prime B - B^{\prime\prime} P^\prime C + P^{\prime\prime}
A^\prime B + B^{\prime\prime} A^\prime P + P^{\prime\prime}
B^\prime C + C^{\prime\prime} B^\prime P \quad ,\cr
F_2 &= - B^{\prime \prime}P^\prime A + C^{\prime\prime}
P^\prime A - A^{\prime\prime} P^\prime B + A^{\prime\prime}
P^\prime C - P^{\prime\prime} A^\prime B - B^{\prime\prime}
A^\prime P - P^{\prime\prime} A^\prime A\cr
&\quad + P^{\prime\prime} B^\prime A - C^{\prime\prime} A^\prime P -
 A^{\prime\prime} A^\prime P + A^{\prime\prime} B^\prime P -
P^{\prime\prime} A^\prime C \quad ,\cr
F_3 &= - \alpha P^{\prime \prime}P^\prime P + B^{\prime\prime}
P^\prime A - C^{\prime\prime} P^\prime A + B^{\prime\prime}
P^\prime C + P^{\prime\prime} A^\prime A - P^{\prime\prime}
B^\prime A + C^{\prime\prime} A^\prime P - P^{\prime\prime}
B^\prime C \quad ,\cr
F_4 &= C^{\prime \prime}P^\prime C - B^{\prime\prime}
P^\prime A + C^{\prime\prime} P^\prime A + C^{\prime\prime}
P^\prime B + B^{\prime\prime} C^\prime P - C^{\prime\prime}
C^\prime P + P^{\prime\prime} C^\prime B\cr
&\quad - P^{\prime\prime} C^\prime C - P^{\prime\prime} A^\prime A +
 P^{\prime\prime} B^\prime A - C^{\prime\prime} A^\prime P -
C^{\prime\prime} B^\prime P \quad ,\cr
F_5 &= - F_3 \quad ,\cr
F_6 &= F_3 \quad ,\cr
F_7 &= \alpha P^{\prime\prime} B^\prime P - 2 \alpha A^{\prime\prime} P^\prime
P - \alpha P^{\prime\prime} P^\prime B - \alpha B^{\prime\prime} P^\prime P
+ B^{\prime\prime} A^\prime C - A^{\prime\prime} A^\prime B -
A^{\prime\prime} B^\prime C \quad ,\cr
F_8 &= - \alpha P^{\prime\prime} B^\prime P - 2 \alpha P^{\prime\prime}
 C^\prime
P + \alpha P^{\prime\prime} P^\prime B + \alpha B^{\prime\prime} P^\prime P
+ B^{\prime\prime} C^\prime C + C^{\prime\prime} C^\prime B -
C^{\prime\prime} B^\prime C \quad ,\cr
F_9 &= \hat F_7 \quad ,\cr
F_{10} &= - \alpha  P^{\prime \prime}B^\prime P + \alpha P^{\prime\prime}
P^\prime B + \alpha B^{\prime\prime} P^\prime P - A^{\prime\prime}
A^\prime A - A^{\prime\prime} B^\prime A - 4 A^{\prime\prime}
C^\prime A\cr
&- B^{\prime \prime}C^\prime A - C^{\prime\prime}
C^\prime A - A^{\prime\prime} C^\prime B - A^{\prime\prime}
C^\prime C + C^{\prime\prime} A^\prime C \quad .\cr}$$
Therefore, the Y--B equation can be satisfied if we have
$$F_1 = F_2 = F_3 = F_4 = F_7 = F_8 = F_{10} = 0 \quad .\eqno(4.7)$$
In contrast to the previous case of $N=8$, we could not, however,
succeed in obtaining
the most general solution of Eq. (4.7).  Consider first the simplest
equation $F_1 = 0$ which can be rewritten as
$$2 \alpha + {C^{\prime\prime} \over P^{\prime\prime}} \
\left( {B^\prime \over P^\prime} - {B \over P} \right) +
{C \over P}\ \left( {B^\prime \over P^\prime} - {B^{\prime\prime} \over
P^{\prime\prime}} \right) + {A^\prime \over P^\prime} \
\left( {B \over P} + {B^{\prime\prime} \over P^{\prime\prime}}
\right) = 0 \quad . \eqno(4.8)$$
Since $\theta^\prime = \theta + \theta^{\prime\prime}$, this gives a
functional equation whose general solutions are diffiuclt to be found in
constrast to Eq. (3.7).  However, we can find some solutions of Eq. (4.8)
as follows.  In analogy to Eq. (3.8), we seek a solution of form
$${B (\theta) \over P(\theta)} = a + b \theta \eqno(4.9)$$
for some constants $a$ and $b$.  Then, Eq. (4.8) can be solved to yield
$$\eqalignno{{C(\theta) \over P(\theta)} &= a^\prime + {d \over b \theta}
\quad ,  &(4.10a)\cr
\noalign{\vskip 3pt}
{A (\theta) \over P(\theta)} &= -a^\prime - {2 (\alpha + d - a a^\prime)
\over b \theta + 2a} &(4.10b)\cr}$$
for some other constants $a^\prime$ and $d$.  Inserting these into the rest
of equations $F_2 = F_3 = F_4 = F_7 = F_8 = F_{10} = 0$,  we can verify
that they are satisfied also, provided that we have
$$a^\prime = a \quad , \quad d = a^2 - \alpha \quad .$$
Therefore, the desired solution is given by
$$\eqalign{{A(\theta) \over P(\theta)} &= - a \quad , \quad
{B (\theta) \over P(\theta)} = a + b \theta \quad ,\cr
\noalign{\vskip 3pt}
{C(\theta) \over P(\theta)} &= a + {a^2 - \alpha \over
b \theta} \cr}\eqno(4.11)$$
where  $a$ and  $b$ are arbitrary constants.  It is interesting to observe
that Eq. (4.11) will be singular--free at $\theta = 0$, if we choose the
arbitrary constant $a$ to satisfy $a^2 = \alpha$.  Its possible relevance
for the knot theory will be  discussed elsewhere.

Also, if we wish, we can find a solution for $P(\theta)= 0$ by first
letting $P(\theta) = {1 \over a}\ F(\theta)$ and $b = ak$ and then taking
the limit $a \rightarrow \infty$ for fixed values of $\alpha, k, \
{\rm and}\ F(\theta)$.  In this way, we obtain a solution of form
$$\eqalign{P(\theta) &= 0 \quad ,\cr
A(\theta) &= - F(\theta) \quad ,\cr
B(\theta) &= \big( 1 + k \theta \big) F(\theta) \quad ,\cr
C(\theta) &= \big( 1 + {1 \over k \theta} \big) F(\theta) \cr}
\eqno(4.12)$$
for an arbitrary function $F(\theta)$ and for arbitrary constant $k$.
This solution corresponds essentially to that of the so(4) model of
 Zamolodchikov's$^{12)}$.
\vfil\eject
\medskip
\noindent {\bf 5. Concluding Remarks}

We have seen in this note that the triple products can be useful for
solving the Yang--Baxter equation.  We may, of course, note contrarily that
we rewrite $R^{ab}_{cd} (\theta)$ first as
$$R^{ab}_{cd} (\theta) = P (\theta) C_{cd}^{\ \ ba} + A(\theta)
g_{cd} g^{ab} + B (\theta) \delta^a_d \delta^b_c +
C(\theta) \delta^a_c \delta^b_d \eqno(5.1)$$
\noindent from Eqs. (1.5), (1.8), (1.12), (1.18) and (2.8), and then determine
$P(\theta),\ A(\theta),\ B(\theta),\ {\rm and}\ C(\theta)$ directly by
inserting Eq. (5.1) into Eq. (1.1a), as has been done in ref. 8.  However,
the use of the triple product makes the calculation far easier.  Moreover,
the direct method becomes hardly manageable for more complicated triple
 systems to be discussed in the subsequent papers.

The triple system discussed in this note is actually a special case of more
general class of system which we call orthogonal triple
 (or ternary) system.  Setting
$$\lambda = -3 \beta \eqno(5.2)$$
and introducing new triple product $xyz$ by
$$xyz = [x,y,z] + \lambda <y \vert z>x - \lambda <z \vert x>y \quad ,
\eqno(5.3)$$
we can easily see that Eqs. (2.1)--(2.5) will lead to the validity of
\item{(i)} $< x \vert y>\  =\  <y \vert x>$ \hfill (5.4)

\item{(ii)} $yxz + xyz = 0$ \hfill (5.5)

\item{(iii)} $xyz + xzy = 2 \lambda <y \vert z> x - \lambda <x \vert y>z -
 \lambda <z \vert x > y$ , \hfill (5.6)

\item{(iv)} $uv(xyz) = (uvx)yz + x(uvy)z + xy(uvz)$\hfill (5.7)

\item{(v)} $<uvx \vert y> = -<x \vert uvy>$ . \hfill (5.8)

\noindent The triple system
 satisfying Eqs. (5.4)--(5.8) is a supersymmetric analogue
of the symplectic triple system$^{13)}$ and may be called the orthogonal
triple system.  Solutions of the Y--B equation in terms of orthogonal or
symplectic triple product systems will be discussed in the proceeding
paper$^{9)}$.  Then, the utilization of mathematical structures of ternary
algebras for these cases becomes essential to find solutions.
\medskip
\noindent {\bf \underbar{Acknowledgement}}

This work is supported in part by the U.S. Department of Energy Grant No.

\noindent DE-FG02-91ER40685.
\bigskip
\bigskip
\bigskip
\noindent {\bf \underbar{Appendix: Properties of Octonionic Triple System}}

We first prove existence of non-trivial solutions of triple systems
satisfying Eqs(2.1)--(2.4) by following the method given elsewhere$^{14)}$.
 Let $L$ be a simple or semi-simple Lie algebra over the complex number
field $F$, and let $V$ be a irreducible module of $L$.  The tensor product
$V \otimes V$ can always be decomposed into a direct  sum of the symmetric
and antisymmetric components, $(V \otimes V)_S$ and
$(V \otimes V)_A$, respectively as
$$V \otimes V = (V \otimes V)_S \oplus (V \otimes V)_A \quad . \eqno(A.1)$$
Analogously, the tensor product $V \otimes V \otimes V$ can be decomposed
into a direct sum of various components of various Young tableau$^{15)}$.
As usual, we specify totally antisymmetric, symmetric and mixed tableaus by
the symbols [1$^3$], [3], and [2,1]
 with $V = [1]$, respectively and so on.  Suppose that
the irreducible module $V$ obeys conditions
$${\rm Dim\ Hom}((V \otimes V)_S \rightarrow F) = 1 \quad , \eqno(A.2)$$
$${\rm Dim\ Hom}([1^3] \rightarrow V) = 1\quad , \eqno(A.3)$$
$${\rm Dim\ Hom}([1^4] \rightarrow F) = 1 \quad , \eqno(A.4)$$
and
$${\rm Dim \ Hom} \{ ([1^3] \otimes [1^3])_S \rightarrow F\} \leq
 2 \quad .\eqno(A.5)$$
Here, Dim $W$ implies the dimension of a vector space $W$, and Hom($W_1
\rightarrow W_2)$ designates the vector space of all homorphisms between
$L$-modules $W_1$ and $W_2$, which commute with actions of $L$.

As we will explain shortly, the validity of Eqs. (A.2)--(A.5)
 implies that of Eqs. (2.1)--(2.4).  Consider first Eq. (A.2) which
implies existence of symmetric bi-linear form $<x \vert y>$.  Moreover, its
invariance under the action of $L$ assures$^{14)}$
 its non-degeneracy, since $V$ is
assumed to be irreducible.  Next, totally antisymmetric triple product
$[x,y,z]$ exists in $V$ in view of Eq. (A.3).  Further, Eq. (2.3)
follows from Eq. (A.4) as we explained in ref. 14.  Finally,
 the validity of Eq.
(A.5) leads to that of Eq. (2.4) for some constants $\alpha$ and $\beta$ by
the same reasoning given there.  Therefore, if we can find a simple Lie
algebra $L$ and its irreducible module $V$ satisfying conditions Eqs.
(A.2)--(A.5), then we have constructed the triple product $[x,y,z]$ in $V$
obeying Eqs. (2.1)--(2.4).  Let $L$ be the Lie algebra $B_3$ which is the
same as the Lie algebra so(7) of the SO(7) group.  Choose $V$ to be its
eight-dimensional spinor representation $V = \{\Lambda_1\}$ so that
$$N = \ {\rm Dim}\ V = 8 \quad . \eqno(A.6)$$
It is easy to verify the validity of Eqs. (A.2)--(A.5) for this case, when
we note for example
$$[1^3] = V \oplus \{ \Lambda_1 + \Lambda_3 \} \quad ,$$
\line{or \hfill (A.7)}
$$56 = 8 + 48$$
where $\Lambda_1,\ \Lambda_2,\ {\rm and}\ \Lambda_3$ are fundamental weights
of $B_3$, and $\{\Lambda_1 + \Lambda_3 \}$ designates the 48-dimensional
irreducible module  of $B_3$ with the highest weight $\Lambda_1 +
\Lambda_3$.  Also, we calculate $[1^4] = \{0\} \oplus \{\Lambda_1\}
\oplus \{2 \Lambda_1\} \oplus \{ 2 \Lambda_3\} \
{\rm or}\ 70 = 1 + 7 +27 + 35$.  Similarly, the same argument applies for
the case of $L= {\rm so}(4)$ with $V$ being its 4-dimensional irreducible
vector representation, although so(4) is not simple but semi-simple.

These two cases discussed above exhaust all possibilities
 for the system satisfying Eqs. (2.1)--(2.5) in some sense to
be specified below.  This fact is also intimately connected with their
relation with octonion and quaternion algebras, as we will show shortly.
First, if $\alpha = 0$, then we must have $\beta =0$ by the result of
section 2 so that we assume $\alpha \not= 0$.  Then, normalizing the triple
product and/or inner product suitably, we may assume hereafter
$$\alpha = 1 \quad . \eqno(A.8)$$
Second, since the inner product $<x \vert y>$ is non-degenerate, there
exists an element $e \ \epsilon\  V$ satisfying $<e \vert e>\ \not= 0$.
Normalizing it suitably, we can set
$$<e \vert e>\ = 1 \quad . \eqno(A.9)$$
For any arbitrary but fixed element $e \ \epsilon\  V$ satisfying Eq. (A.9),
we can now introduce a bi-linear product
$$x \cdot y \ ;\  V \otimes V \rightarrow V \eqno(A.10)$$
by
$$x \cdot y = [x, y,e]\  + <x \vert e>y\  + <y \vert e>x\  -
<x \vert y> e \quad . \eqno(A.11)$$
Here, we used the symbol $x \cdot y$ rather than the customary $xy$ in
order to avoid possible confusions with the triple product $xyz$ defined by
Eq. (5.3).  Moreover, we define the conjugate of $x$ as usual by
$$\overline x = 2<x \vert e>e - x \quad . \eqno(A.12)$$
It is easy to verify from
Eqs. (2.1)--(2.5) with $\alpha = 1$ that this bi-linear product satisfies the
composition law
$$<x \cdot y \vert x \cdot y> \ =\  < x \vert x ><y \vert y> \eqno(A.13)$$
as well as
$$\eqalign{x \cdot e &= e \cdot x = x \quad ,\cr
x \cdot \overline x &= \overline x \cdot x =\  <x \vert x >e \quad ,\cr
x \cdot (\overline x \cdot y) &= (x \cdot \overline x)\cdot y =\
<x \vert x > y \quad ,\cr
(y \cdot x)\cdot \overline x &= y \cdot (x \cdot \overline x) =
\ <x \vert x> y \quad , \cr
<x \cdot y \vert z> &=\  < \overline x \cdot z \vert y >\quad .\cr}
\eqno(A.14)$$
In other words, it defines a quadratic alternative composition
algebra$^{16)}$, so that possible dimensions of $V$ are limited to be $N=
 1,2,4\ {\rm or}\ 8$ by the Hurwitz's theorem$^{16)}$.  Moreover, since two
cases of $N=1$, and 2 lead to the trivial case $[x,y,z] = 0$ identically,
this reproduces the result of section 2.  Two cases of $N=4$ and 8
correspond to the quaternion and octonion algebras.  First consider the
quaternion case of $N=4$.  Let $e(= e_4),\ e_1,\ e_2,\ {\rm and}\ e_3$ be
the standard quaternionic basis with normalization
$$< e_\mu \vert e_\nu> \ = \delta_{\mu \nu}$$
for $\mu, \nu = 1,2,3,4$.  Then, we can readily show the validity of
$$<e_\alpha \vert [ e_\mu, e_\nu, e_\lambda]> \ = \epsilon_{\alpha
 \mu \nu \lambda}$$
where $\epsilon_{\alpha \mu \nu \lambda} \ {\rm with}\
\epsilon_{1234} = 1$ is the totally antisymmetric Levi-Civita
symbol in the 4-~dimensional space.  This, then, uniquely determines the
triple product for $N=4$ by Eq. (2.10).

For $N=8$, Eq. (2.6a) with Eq. (A.8) determines $\beta$ to be $\pm 1$.
Changing the sign of the triple product, if necessary, we can then assume $
\beta = -1$.  Setting $z = v = e$ in Eq. (2.5), and then changing $u$ into
$z$, it gives
$$(x \cdot y)\cdot z = [x,y,z] + 2<y \vert e>x \cdot z
\  - <y \vert z>x\  - <x \vert y>z\  + <x \vert z>y \quad . \eqno(A.15)$$
Antisymmetrizing Eq. (A.15) together with Eq. (A.14), we can rewrite Eq.
 (A.15) into a more symmetrical form of
$$\eqalign{[x,y,z] =
 &{1 \over 2}\ \{ (x,y,z)\  + <x \vert e>[y,z]\  + <y \vert e> [z,x]\cr
 &+ <z \vert e> [x,y] - <z \vert [x,y]>e\}\cr}\eqno(A.16)$$
where $(x,y,z)$ and $[x,y]$ are associator and commutator, respectively,
defined by
$$\eqalign{(x,y,z) &= (x \cdot y)\cdot z - x \cdot (y \cdot z)\cr
[x,y] &= x \cdot y - y \cdot x \quad .\cr}\eqno(A.17)$$
We may mention the fact that in derivation of Eq. (A.16), we used the
identity$^{17)}$
$$\eqalign{[[x,y],z] &+ [[y,z],x] + [[z,x],y]\cr
&= (x,y,z) + (y,z,x) + (z,x,y) - (z,y,x) -
(y,x,z) - (x,z,y)\cr}\eqno(A.18)$$
which is equal to $6(x,y,z)$ for alternative algebra.  Since the triple
product $[x,y,z]$ can be determined uniquely in terms of the octonion
algebras by Eq. (A.15) or (A.16), we may say  that the eight
dimensional triple system is also unique.  These are the reasons also why
we called our triple products defined by Eqs. (2.1)--(2.4) to be
quaternionic and octonionic ternary systems, respectively, for $N=4$ and 8.
 In spite of relationship between the octonion algebra and the octonionic
triple system, the triple system enjoys a larger symmetry so(7) by
construction in comparison to $G_2$ of the octonion.  The reason is, of
course, due to the introduction of the fixed privileged element $e$, which
breaks$^{18)}$ the symmetry from the so(7) to $G_2$. Its relation to
so(8) will be discussed in the proceeding paper$^{9)}$.

In this connection, we remark that Eq. (A.16) can be generalizeable for any
orthogonal triple system defined in section 5 by introducing the
antisymmetric product $[x,y]$ by
$${1 \over 2}\ [x,y] = [x,y,e] = xye + \lambda
<x \vert e>y - \lambda <y \vert e>x \eqno(A.19)$$
for any privileged element $e\ \epsilon\ V$ satisfying $<e \vert e>\  = 1$.
Then, the relations Eqs. (5.4)--(5.8) can be used to prove the validity of
the identity
$$\eqalign{4 \lambda [x,y,z] = &[[x,y],z] +
[[y,z],x] + [[z,x],y]\cr
&+ 2 \lambda \{ <e \vert x> [y,z]\  + <e \vert y>[z,x]\cr
& +\  <e \vert z>[x,y]\  - <z \vert [x,y]>e\}\cr} \eqno(A.20)$$
by setting $v = z = e$ in Eq. (5.7) and then rewriting $u$ by $z$.  If
$\lambda \not= 0$, it implies that the triple product $[x,y,z]$ can be
expressed in terms of the bi-linear product $[x,y]$.  Indeed, for the
octonion algebra, we have $\lambda = -3 \beta = 3$ and Eq. (A.20) will
immediately reproduce Eq. (A.16).  However for the case $\lambda = 0$ as in
the quaternionic triple system, Eq. (A.20) gives only
$$[[x,y],z] + [[y,z],x] + [[z,x],y] = 0 \eqno(A.21)$$
so that the bi-linear product $[x,y]$ must be automatically a Lie algebra.

Next, we would like to mention a fact that we can construct another type
of a triple product from the octonion algebra, following the works of
Allison$^{19)}$ and Kamiya$^{20)}$.  Defining the left and right
multiplication operators $L_x$ and $R_y$, respectively, by
$$L_x y = x \cdot y \quad , \quad R_x y = y \cdot x \quad, \eqno(A.22)$$
we know$^{16)}$ that
$$D_{x,y} = [R_x , R_y] + [L_x, L_y] + [L_x , R_y]\eqno(A.23)$$
is its derivation, i.e., it satisfies
$$D_{x,y} (u \cdot v) = (D_{x,y} u) \cdot v + u \cdot
(D_{x,y} v) \quad . \eqno(A.24)$$
We, now, define a new triple product $x*y*z$ by
$$\eqalign{x*y*z &= - {1 \over 4} \ D_{x,y} z\cr
\noalign{\vskip 3pt}
&= {1 \over 4}\ \{y \cdot (x \cdot z) - (z \cdot y) \cdot x +
 (z \cdot x + x \cdot z)\cdot y - x \cdot
(y \cdot z+ z \cdot y)\} \quad .\cr}\eqno(A.25)$$
Since $D_{x,y}$ is a derivation of any alternative algebra, we have
also
$$\eqalign{D_{u,v}(x*y*z) = &(D_{u,v}x)*y*z + x*(D_{u,v}y)*z\cr
&+ x*y*(D_{u,v}z)\cr}$$
because of Eqs. (A.24) and (A.25).  In other words, we have
$$\eqalign{u*v&*(x*y*z)\cr
&= (u*v*x)*y*z + x*(u*v*y)*z + x*y*(u*v*z) \quad .\cr}\eqno(A.26)$$
Using Eqs. (A.14) and (A.18), we can rewrite Eq. (A.25) to be
$$\eqalign{x*y*z &= {3 \over 4}\ (x,y,z) - {1 \over 4}\ [[x,y],z]\cr
\noalign{\vskip 3pt}
&= {1 \over 4}\ (x,y,z) + \{<y \vert z> \ - \ <y \vert e><z \vert
e>\}x\cr
\noalign{\vskip 3pt}
&\quad - \{<x \vert z> \ -\ <x \vert e><z \vert e>\}y\cr
\noalign{\vskip 3pt}
&\quad - \{ <x \vert e><y \vert z> \ -\ <y \vert e><x \vert z>\}e\cr}
\eqno(A.27)$$
\noindent from which we find
\itemitem{(i)} $x*y*z + y*x*z = 0$

\itemitem {(ii)} $x*y*z + x*z*y$
\itemitem {    } $= \{2<y \vert z>\ -\ 2<y \vert e><z \vert e>\}x$
\itemitem {    } $\quad - \{ <x \vert z>\ -\ <x \vert e><z \vert e>\}y$
\hfill (A.28)
\itemitem {    } $\quad - \{ <x \vert y>\ -\ <x \vert e><y \vert
e>\}z$
\itemitem {    } $\quad + \{<y \vert e><x \vert z> \ +
\ <z \vert e><x \vert y>\ -
\ 2<y \vert z><x \vert e>\}e \quad .$

\noindent The relationship between $x*y*z$ and $xyz$ given in section 5
 is not straightforward.
  First, for $N=4$, we find for instance the Lie--triple condition
$$[x,y,z]^* = {1 \over 3}\ \{ x*y*z + y*z*x + z*x*y\} = 0$$
while the quaternionic triple product
 $[x,y,z]$ does not satisfy the corresponding
relation.  For $N=8$, on the other side, we obtain
$$\eqalign{2x*y&*z -xyz\cr
&= \{<x \vert z> \ +\  2< x \vert e><z \vert e>\}y - \{<y \vert z>\ +\
2 <y \vert e><z \vert e>\}x\cr
&\quad - 2\{<x \vert e><y \vert z>\ -\ <y \vert e><x \vert z>\}e -
 {1 \over 2}\ \{<x \vert e>[y,z] \cr
&\quad +\ <y \vert e>[z,x]\  + \ <z \vert e>[x,y]\  - \
<z \vert [x,y]>e\} \quad .\cr}\eqno(A.29)$$
\noindent From Eqs. (A.26) and (A.28), we see that the new triple product
$x*y*z$ satisfies almost the same relations
 as Eqs. (5.4)--(5.7).  We can make
it to be the orthogonal ternary system, if we  restrict ourselves to the
 sub-space
$$V_0 = \{x \vert <x \vert e>\  = 0 \quad , \quad x\ \epsilon\ V\}
\eqno(A.30)$$
so that $N_0 = \ {\rm Dim}\ V_0 = 3$ or 7 according to $N=4$ or 8.  Since
we can readily verify
$$<e \vert x*y*z>\  = 0 \eqno(A.31)$$
whenever we have $x,y,z\ \epsilon\ V_0$, the product $x*y*z$ defines a
 triple product in $V_0$.
 Moreover, Eqs. (A.26) and (A.28) show now that it satisfies all
axioms of orthogonal ternary systems with $\lambda$ now being
1 for $V_0$.  However for $N_0 = 3$, it is trivial in a sense that we have
$$x*y*z = \ <y \vert z>x\  - \ <x \vert z>y \quad .$$
For $N_0 =7$, the space $V_0$
 corresponds to the 7-dimensional
simple exceptional Malcev algebra$^{17)}$, so that we call the present
system as the Malcev triple system.  Although we have introduced
$x*y*z$ in terms of quadratic alternative algebra, we could have
started directly with the known derivations$^{21)}$ of any Malcev
algebra.  However we will find the same result for $N_0 = 7$ as is expected.

We can characterize the orthogonal triple system just as we have done for
both quaternionic and octonionic triple system by Eqs. (A.2)--(A.5).
Suppose that we now have
$$\eqalign{&{\rm Dim\ Hom}\ ((V \otimes V)_S \rightarrow F) = 1
\quad ,\cr
&{\rm Dim\ Hom}\ ([1^3] \rightarrow V) = 1
\quad ,\cr
&{\rm Dim\ Hom}\ ([1^4] \rightarrow F) = 1
\quad ,\cr}\eqno(A.32)$$
just as Eqs. (A.2)--(A.4), but we now assume
$${\rm Dim\ Hom}\ ([1^4]  \otimes [1^2] \rightarrow F) \leq 1 \eqno(A.33)$$
instead of Eq. (A.5).  Then, following the method given in ref. 14, the
totally antisymmetric ternary product $[x,y,z]$ can be shown to satisfy
$$\eqalign{[u,v,[x,y,z]] &- [[u,v,x],y,z] - [x,[u,v,y],z]
 - [x,y,[u,v,z]]\cr
&= \lambda \{ <u\vert [x,y,z]>v\  - <v \vert [x,y,z]>u\  + \
<x \vert v>[u,y,z]\cr
&\quad  + <y \vert v>[x,u,z] \
 +\ <z \vert v>[x,y,u]\  -
\ <x \vert u>[v,y,z]\cr
&\quad -\ <y \vert u>[x,v,z]\  -\
<z \vert u>[x,y,v]\}\cr}
\eqno(A.34)$$
for a constant $\lambda$.  Setting
$$xyz = [x,y,z] + \lambda \{<y \vert z>x\  - \
<z \vert x>y \} \quad , \eqno(A.35)$$
it can be shown to satisfy the axioms Eqs. (5.3)--(5.8) of the orthogonal
triple system.  If we choose $V= \{\Lambda_2 \}$ to be the 7-dimensional
irreducible module of the exceptional Lie algebra $G_2$, then we can verify
the validity of Eqs. (A.32) and (A.33) so that it reproduces the same
triple product $x*y*z$, where we have used for example
$$\eqalign{[1^2] &= \{ \Lambda_2 \} \oplus \{\Lambda_1 \} \quad
{\rm or}\quad 21 = 7 + 14 \quad , \cr
[1^3] &= [1^4] = \{0 \} \oplus \{ \Lambda_2 \} \oplus
\{ 2\Lambda_1 \} \quad {\rm or}\quad 35 = 1 + 7 + 27 \quad .
\cr}\eqno(A.36)$$
Moreover, we can verify
$${\rm Dim\ Hom}\ ([2,1] \rightarrow V) = 1 \eqno(A.37)$$
since we find
$$[2,1] = \{ \Lambda_2 \} \oplus \{ \Lambda_1 \} \oplus \{
2 \Lambda_2 \} \oplus \{ \Lambda_1 + \Lambda_2 \}$$
\line{or \hfill (A.38)}
$$112 = 7 + 14 + 27 + 64 \quad .$$
Conditions Eqs. (A.32), (A.33) and (A.37) also hold  valid for $V$ being
 8-dimensional spinor representation $\{ \Lambda_3 \}$ of so(7) = $B_3$, as
we expect.  The condition Eq. (A.37) will
 turn out to be crucial for the calculation of
the proceeding paper for both cases.

In ending this note, we  simply remark that there exists$^{22)}$ another
entirely different class of triple systems which are nevertheless
intimately related also to the octonion algebra.  However, we will not go
into detail.

\vfil\eject
\noindent {\bf \underbar{References}}
\item{1.} \lq\lq Yang--Baxter Equation in Integrable Systems", ed. by M.
Jimbo, World Scientific, Singapore (1989).
\item{2.} \lq\lq Integrable Systems and Quantum Groups", ed. by M. Carfora,
M. Martelli, and A. Marzuoli, World Scientific, Singapore (1992).
\item{3.} \lq\lq Braid Group, Knot Theory and Statistical Mechanics", ed.
by C. N. Yang and M. L. Ge, World Scientific, Singapore, (1989).
\item{4.} L. H. Kauffman,\lq\lq Knots and Physics", World Scientific,
Singapore, (1991).
\item{5.} Y. I. Manin, \lq\lq Quantum Groups and Non-Commutative Geometry",
University of Montr\'eal Press, Montr\'eal, (1988).
\item{6.} K. Meyberg, \lq\lq Lectures on Algebras and Triple Systems"
 Lecture Note, the University of Virginia, Charlottesville (1972)
 (unpublished).
\item{7.} \lq\lq On Triple Systems", ed. by K. Yamaguchi, Research
Institute for Mathematical Sciences, Kyoto University (1977) (unpublished
mimeograph in Japanese).
\item{8.} H. J. de Vega and H. Nicolai, Phys. Lett. {\bf B244}, 295
(1990).
\item{9.} S. Okubo, University of Rochester Report, UR--1294 (1992).
\item{10.} B. de Wit and H. Nicolai, Nucl. Phys. {\bf B231}, 506 (1984).
\item{11.} F. G\"ursey and C. H. Tze,
  Phys. Lett. {\bf 127B}, 191 (1983).
\item{12.} A. B. Zamolodchikov and A$\ell$. B. Zamolodchikov,
Nucl. Phys. {\bf B133}, 525
 (1978).
\item{13.} K. Yamaguchi and H. Asano, Proc. Jap. Acad. {\bf 51}, 247
(1972).
\item{14.} S. Okubo, Alg. Group. Geom. {\bf 3}, 60 (1986).
\item{15.} H. Weyl, \lq\lq Classical Group", Princeton University Press,
 Princeton, NJ (1939).
\item{   } D. E. Littlewood, \lq\lq The Theory of Group
 Character", Clarendon, Oxford (1940).
\item{16.} R. D. Schafer, \lq\lq An Introduction to Non-associative
Algebras", Academic Press, New York/London (1966).
\item{17.} H. C. Myung, \lq\lq Malcev--Admissible Algebras", Birkh\"auser,
Boston, MA (1986).
\item{18.} R. E. Behrends, J. Dreitlein, C. Fronsdal, and W. Lee, Rev. Mod.
Phys. {\bf 34}, 1 (1962).
\item{19.} B. N. Allison, Math. Ann. {\bf 237}, 133 (1978), J. Alg.
 {\bf 143}, 63 (1991).
\item{20.} N. Kamiya, Preprint, Shimane University,
 Matsue, Japan (1991) to appear in Alg. Group. Geom.
\item{21.} A. A. Sagle, Trans.. Amer. Math. Soc. {\bf 101}, 426 (1961).
\item{22.} S. Okubo, Hadronic Jour. {\bf 1}, 1383, 1432 (1978), {\bf
 2}, 39 (1979).

\bye